\documentclass[letterpaper,10pt]{article} 

\usepackage{opticameet3} 

\newcommand\authormark[1]{\textsuperscript{#1}}

\usepackage{amsmath,amssymb}
\usepackage[colorlinks=true,bookmarks=false,citecolor=blue,urlcolor=blue]{hyperref} 

\begin{document}

\title{A comment on Guo et al. (2022)}


\author{Ben Lonnqvist,\authormark{1,*} Harshitha Machiraju,\authormark{1, 2} and Michael H. Herzog\authormark{1}}

\address{\authormark{1} Laboratory of Psychophysics \\
\authormark{2}Signal Processing Laboratory\\
\authormark{}École Polytechnique Fédérale de Lausanne (EPFL), Switzerland}

\email{\authormark{*}benhlonnqvist@gmail.com} 



In a recent article, \textcite{guo2022} report that adversarially trained neural representations in deep networks may already be as robust as corresponding primate IT neural representations. 
By careful perturbation, the authors could change the preferred viewing image of IT neurons—--for example, a neuron that had the highest firing rate for images of pressure gauges responded most strongly to images of dogs when minor perturbation to the image was applied. 
The authors report that the degree of image perturbation required to do this may be even lower than for adversarially trained deep networks. 
The authors interpret this result by posing an apparent paradox: ‘How is it that primate visual perception seems so robust yet its fundamental units of computation are far more sensitive than expected?’

While we find the paper’s primary experiment illuminating, we have doubts about the interpretation and phrasing of the results presented in the paper.

The postulated paradox only seems paradoxical because one of the authors' premises does not hold true---namely that IT is sparsely coding for object category. 
Indeed, if the monkey was not behaviorally fooled by the neuron-specific adversarial perturbations, as we assume, the data rather show that IT is either not coding for object category, or it is not \textit{sparsely} coding for object category. 
Since, as acknowledged by the authors, IT is known to demonstrate object category selectivity \parencite[e.g.][]{gross_visual_1972, kanwisher_fusiform_1997, bao_map_2020}, the data show that object category is not sparsely coded in IT. 
The assumption behind the interpretation of the results is demonstrated false by the results themselves.

In general, it is neither a necessary nor a sufficient condition for any neuron in a system to be robust to adversarial attacks for the system to be adversarially robust \parencite[e.g.][]{gavrikov_adversarial_2022, pauli_training_2022, kumari_harnessing_2019}. 
This is due to the compositional functional structure of systems with hierarchical structure, such as deep learning systems and the primate visual system. 
In other words, the results about IT neurons not being adversarially robust do not imply anything about the robustness of primate visual perception as a whole.

That being said, we think that these findings are timely and exceptionally important as they highlight problems with the common method of assessing deep network models of vision on the basis of their representational similarity\footnote{Or other neurally based metrics}  to neural recordings in primate cortex. 
If cortex codes in a population-wide manner, recording small populations of neurons in primate cortex appears to give few guarantees of capturing a substantially informative part of this population-wide code.

In summary, the results shown in the paper do not demonstrate evidence for the lack of robustness of the primate visual stream, but rather call for increased focus to be placed on behaviorally based metrics of deep network modelling of the visual system \parencite[e.g.][]{elsayed_adversarial_2018, dujmovic_what_2020, geirhos_partial_2021}. 
In addition, we believe that the paper demonstrates that larger-scale neural recordings and composite datasets of many small-scale neural recordings \parencite[e.g.][]{Schrimpf2020integrative} are crucially needed to improve the quality of model judgments based on primate neural recordings.

\printbibliography

\end{document}